\documentclass[runningheads]{llncs}

\usepackage{subcaption}
\usepackage{url}
\usepackage[T1]{fontenc}
\usepackage{graphicx}
\usepackage{hyperref}
\usepackage{psfrag}
\usepackage[ruled,vlined,linesnumbered,algosection,resetcount]{algorithm2e}
\usepackage{amsmath}
\usepackage{amsfonts}
\usepackage{amssymb}
\usepackage{amsbsy}
\usepackage{mathrsfs}
\usepackage{a4}
\usepackage{tabularx}
\usepackage{xcolor}
\usepackage{multirow}
\usepackage{float}
\usepackage{upgreek}
\usepackage{dsfont}
\usepackage[noend]{algorithmic}
\algsetup{indent=2em}
\usepackage{booktabs}
\usepackage{multirow}
\usepackage{cleveref}
\DeclareMathOperator*{\argmin}{argmin}

\newcommand{\bb}[1]{\mathbb{#1}}
\renewcommand{\v}[1]{\boldsymbol{#1}}
\newcommand{\m}[1]{\mathbf{#1}}

\renewcommand{\c}[1]{\mathcal{#1}}

\newcommand{\gvn}{\,|\,}

\newcommand{\var}{\mathbb{V}\mathrm{ar}}
\newcommand{\X}{{\color{blue}\m X}}

\newcommand{\lt}{\left}
\newcommand{\rt}{\right}
\renewcommand{\upupsilon}{\eta}
\newcolumntype{C}{>{\centering\arraybackslash}X}

\def\boxit#1{\vbox{\hrule\hbox{\vrule\kern6pt
          \vbox{\kern6pt#1\kern6pt}\kern6pt\vrule}\hrule}}

\def\boxit#1{\vbox{\hrule\hbox{\vrule\kern6pt
          \vbox{\kern6pt#1\kern6pt}\kern6pt\vrule}\hrule}}

\def\boxit#1{\vbox{\hrule\hbox{\vrule\kern6pt
          \vbox{\kern6pt#1\kern6pt}\kern6pt\vrule}\hrule}}

\begin{document}
\title{Feature Selection in Generalized Linear models via  the Lasso: To Scale or Not to Scale?}
%
%
\author{Anant Mathur\inst{1} \and
Sarat Moka\inst{1}\and Zdravko Botev\inst{1}}
\authorrunning{Mathur et al.}
\titlerunning{Generalized Linear models via  the Lasso: To Scale or Not to Scale?}%
%
\institute{University of New South Wales,
Kensington,
Sydney, NSW 2052, Australia, \email{anant.mathur@unsw.edu.au}
\\
}
\maketitle              
\begin{abstract}
The Lasso regression is a popular regularization method for feature selection in statistics. 
Prior to computing the Lasso  estimator in both linear and generalized linear models, it is common to conduct a preliminary rescaling of the feature matrix to ensure that all the features are standardized.  Without this standardization, it is argued, the Lasso estimate will unfortunately depend on the units used to measure the features. We propose a new type of iterative rescaling of the features in the context of generalized linear models. Whilst existing Lasso algorithms perform a single scaling as a preprocessing step,  the proposed rescaling is applied iteratively throughout the Lasso computation until convergence.  We provide numerical examples, with both real and simulated data, illustrating that the proposed iterative rescaling can significantly improve the statistical
 performance of the Lasso estimator without incurring any significant  additional computational cost. 

\end{abstract}

\section{Introduction and Background}

We begin with providing the background  notation and facts for  the  linear  regression model and its extensions to generalized linear models. With this requisite  notation and background material, we will then be able to explain the main contribution of our paper without unnecessary verbosity.

  We denote the $n\times p$  regression matrix (or feature matrix) containing  the  $p$ features $\v v_1,\ldots,\v v_p$  as 
\[
\X=[\v v_1,\ldots,\v v_p]=\begin{bmatrix}
	\v x_1^\top\\
	\vdots\\
	\v x_n^\top
	\end{bmatrix},
\] 
and the corresponding regression response vector as $\v Y\in\bb R^n$ (with $\v y$ being the realization of the random vector $\v Y$). We make the standard assumptions (see, for example, \cite{kroese2019data})  that  the regression function, or the mean of $\v Y$ conditional on $\X$, satisfies $\v Y=\bb E_{\X}[\v Y]+\v\epsilon$ for some noise vector $\v\epsilon$ with conditional expectation $\bb E_\X[\v\epsilon]=\v 0$ and conditional variance $\var_\X(\v\epsilon)=\sigma^2\m I_n$,  where $\sigma$ is an unknown parameter. Recall that in a linear model, we assume that 
\[
\bb E_{\X}[\v Y]=\v 1\beta_0+\X\v \beta
\]
is a linear function of some model coefficients $\v\beta\in \bb R^p$ and $\beta_0\in \bb R$, the last one being  called the intercept and corresponding to the constant feature  $\v 1\in \bb R^n$. Thus, the   model for $\v Y$ is 
\[
\v Y=\v 1\beta_0+\X\v \beta+\v\epsilon.
\]
Define the projection (idempotent) matrix $\m C:=\m I_n-\v 1\v 1^\top/n$ and let
\[
\eta_i^2:=\frac{\|\m C\v v_i\|^2}{n}=\frac{\v v_i^\top\m C \v v_i}{n},\quad i=1,\ldots,p
\]
be the empirical variance of the components of the $i$-th feature vector.
We define the Lasso estimate \cite{tibshirani1996regression} of $\v\beta$ as the solution to the penalized least squares:
\begin{equation}
	\label{lasso}
	(\hat\beta_{0,\lambda},\hat{\v \beta}_\lambda)=\argmin_{b_0,\v b}\frac{\|\v y-\v 1 b_0-\X\v b\|^2}{2n}+\lambda\sum_{i=1}^p\upupsilon_i\times|b_i|,
\end{equation}
where $\lambda>0$ is a suitably chosen regularization parameter and the intercept $b_0$ is not penalized. 
It will shortly become clear  why the definition \eqref{lasso} of the lasso estimator appears prima facie to be different from the  one used in  textbooks \cite{hastie2015statistical} and statistical packages \cite{friedman2010regularization}.
We remind the reader that the main advantage of the Lasso  regularization is that many of the components of 
$\hat{\v \beta}_\lambda$ are estimated as zeros, making it very easy to tell which features are important and which are not. This phenomenon is behind the well-known ability of the Lasso estimator to perform simultaneous shrinkage and model selection \cite{hastie2015statistical,tibshirani1996regression}.

\paragraph{Feature Standardization.}
As mentioned in the abstract, it is common practice to standardize the  features $\v v_1,\ldots,\v v_p$ so that the variance of each $\v v_i$  is unity \cite{hastie2015statistical,tibshirani1996regression}. 
In other words,  the columns $\v v_1,\ldots,\v v_p$ of 
$\X$ are all rescaled in such a way that $\|\m C\v v_i\|^2=n$ for all $i$. This standardization ensures that the Lasso estimate $\hat{\v\beta}_\lambda$ is not affected by the units in which the features are measured, and in general improves the performance of the estimator \cite{hastie2009elements}. 
The standardization can be accomplished by working with the matrix $\X\m \Upsilon$, rather than $\X$, where $\m \Upsilon$ is the rescaling matrix  
\[
\m \Upsilon:=\mathrm{diag}(\upupsilon_1^{-1},\ldots,\upupsilon_p^{-1}).
\]
The solution to \eqref{lasso} (without preliminary standardization of the features) is equivalent to the solution
\begin{equation}
	\label{lasso standard}
	(\hat\beta_{0,\lambda},\m \Upsilon^{-1}\hat{\v \beta}_\lambda)=\argmin_{b_0,\v b}\frac{\|\v y-\v 1 b_0 -[\X\m \Upsilon]\v b\|^2}{2n}+\lambda\sum_{i=1}^p|b_i|,
\end{equation}
so that  $\hat{\v \beta}_\lambda$ is now in agreement with the definition of the Lasso estimator given  in  textbooks and statistical packages \cite{friedman2010regularization}. 

\paragraph{Extensions to GLMs.} 
Suppose that the joint density of the response vector $\v Y$ given $\beta_0,\v\beta,\X$ is $g(\v y\gvn \beta_0,\v\beta,\X )$,  where
the dependence on $\beta_0,\v\beta,\X$ is through the linear map $(\beta_0,\v\beta)\mapsto\v 1\beta_0+\X\v\beta$.
Here, the cross-entropy training loss \cite{kroese2019data} (negative average log-likelihood) is $-\frac{1}{n}\ln g(\v y\gvn \beta_0,\v\beta,\X )$ and
 the extension of the Lasso estimator \eqref{lasso standard} to the setting of generalized linear models is then given by :
\begin{equation}
	\label{glm common}
	(\hat\beta_{0,\lambda},\m\Upsilon^{-1}\hat{\v \beta}_\lambda)=\argmin_{b_0,\v b}\frac{-\ln g(\v y\gvn b_0, \v b,\X\m \Upsilon )}{n}+\lambda\sum_{i=1}^p|b_i|.
\end{equation}
Observe that, just like in the linear Lasso estimator \eqref{lasso standard}, we scale the features so that their variance is unity \cite{hastie2015statistical}.  This scaling need only be applied once on $\X$, possibly as a preprocessing step prior to the main optimization, and then reversed at the end of the optimization  (in order to obtain the regression coefficients in the original units of measurement). The Lasso solution \eqref{glm common} can be also be rewritten in an equivalent  form to  \eqref{lasso}, namely, 
\begin{equation}
	\label{glm scaled penalty}
	(\hat\beta_{0,\lambda},\hat{\v \beta}_\lambda)=\argmin_{b_0,\v b}\frac{-\ln g(\v y\gvn b_0,\v b,\X )}{n}+\lambda\sum_{i=1}^p\upupsilon_i\times|b_i|.
\end{equation}
We are now ready to describe our proposed methodology of rescaling.
\paragraph{New Rescaling Method for GLMs.} 
Let
\[
r(b_0,\v b):=\frac{-\ln g(\v y\gvn b_0, \v b,\X)}{n}
\]
be our shorthand notation for the cross-entropy loss.
We define 
\[
\upupsilon_i^2(\v\beta),\qquad i=1,\ldots,p,
\] 
to be the $i$-th diagonal element of the $p\times p$ Hessian matrix of second derivatives:
\[
\frac{\partial^2}{\partial\v\beta\partial\v\beta^\top}\min_{b_0}r(b_0,\v \beta).
\]
This is the Hessian matrix of the cross-entropy loss, evaluated at the true parameter $\v\beta$, and after  the nuisance parameter $\beta_0$ is eliminated from the optimization.
Then, instead of the usual rescaled Lasso estimator \eqref{glm scaled penalty}, in this article we propose the following alternative \emph{iteratively rescaled Lasso} (IRL):
\begin{equation}
	\label{glm new}
 \argmin_{\v b,b_0}\frac{-\ln g(\v y\gvn b_0,\v b,\X )}{n}+\lambda\sum_{i=1}^p\upupsilon_i(\v\beta)\times|b_i|.
\end{equation}
We now  make three observations.

First, since each $\upupsilon_i(\v\beta)$ depends on the unknown $\v\beta$, the approximate computation of \eqref{glm new} will be iterative --- this will be  explained carefully in the next section --- and is the main reason for naming the method IRL. 

Second, the  linear Lasso estimator \eqref{lasso} is a special case of \eqref{glm new}   when
$\v Y$ is a multivariate Gaussian with mean $\bb E_\X[\v Y]=\v 1\beta_0+\X\v\beta$ and variance $\var_\X(\v Y)=\m I$, because then $\upupsilon_i^2(\v\beta)=\|\m C\v v_i\|^2/n$.

Third, one may ask what is the motivation for the proposed IRL estimator.  The answer is that the IRL estimator  coincides with the traditional linear regression estimator \eqref{lasso}, provided that the cross-entropy loss $r(b_0,\v b)$ is replaced by its quadratic approximation in the neighborhood of the true coefficients $\beta_0,\v\beta$.  In other words, our proposed IRL estimator uses exactly the same scaling as the linear Lasso estimator \eqref{lasso} when the generalized linear model is \emph{linearized} at the true solution. Note that there is no such agreement in the scaling between the currently accepted linear estimator \eqref{lasso} and its GLM counterpart \eqref{glm scaled penalty}, that is, the current widely-used scaling is not consistent across  linear and nonlinear models.  Our proposal is thus motivated by the desire for consistency in the scaling applied to linear and nonlinear models.

The rest of the paper is organized as follows. 
In Section~\ref{sec:compute} we explain how to approximately compute the Lasso estimator in
\eqref{glm new}, given that we do not actually have apriori knowledge of the Hessian matrix $\m H$ at the true parameter $\v\beta$. Then, in Section~\ref{sec: numerical} we provide a number of numerical examples with both simulated and real data illustrating the scope of  improvement in the statistical performance of the Lasso estimator. In the final section, we give concluding remarks.

\section{Computation via Iterative Reweighted Least Squares}
\label{sec:compute}
Since the true $\v\beta$ is not known apriori, in this section we explain how to approximately compute \eqref{glm new} using an iterative reweighted least squares (IRLS) method. 
We begin by  reviewing 
 the well-known IRLS for computing the estimator \eqref{glm scaled penalty} and then explain how it is easily modified to approximately compute our proposed estimator \eqref{glm new}. 
To this end, we introduce the notation  $\breve{\v b}:=[b_0,\v b^\top]^\top$ and $\breve{\X}:=[\v 1,\X]$, so that
$
\breve{\X}\breve{\v b}=\v 1 b_0+\X\v b.
$
 Then, computing \eqref{glm scaled penalty} is equivalent to minimizing 
 \[
 r(\breve{\v b})+\lambda\sum_{i=1}^p \upupsilon_i\times |b_i|
 \]
 with respect to $\breve{\v b}$.
 This problem is nonlinear, but it  can be solved by successive and repeated linearizations of $r(\breve{\v b})$. 
 Suppose that at iteration $t$, we have a current best guess $\breve{\v b}_t$ for the  minimizer of \eqref{glm scaled penalty}. 
 Given this $\breve{\v b}_t$, we consider the quadratic multivariate Taylor approximation to  the cross-entropy loss at the point $\breve{\v b}_t$:
 \begin{equation*}
 	r(\breve{\v b}_t)+ (\breve{\v b}-\breve{\v b}_t)^\top\nabla r(\breve{\v b}_t)+\frac{1}{2} (\breve{\v b}-\breve{\v b}_t)^\top\m H(\breve{\v b}_t)(\breve{\v b}-\breve{\v b}_t).
 \end{equation*}
 Then, we update $\breve{\v b}_t$ to $\breve{\v b}_{t+1}$ by computing the linear Lasso estimator:
 \begin{equation}
 	\label{Taylor}
 \breve{\v b}_{t+1}:=\argmin_{\breve{\v b}}\;\;(\breve{\v b}-\breve{\v b}_t)^\top\nabla r(\breve{\v b}_t)+\frac{1}{2} (\breve{\v b}-\breve{\v b}_t)^\top\m H(\breve{\v b}_t)(\breve{\v b}-\breve{\v b}_t)+\lambda\sum_{i=1}^p \upupsilon_i\times|b_i|.
 \end{equation}
 This computation is then iterated until convergence \cite{friedman2010regularization}.

To keep the mathematical detail simple, we henceforth use the Logit model to illustrate the computations that are typically required for all GLMs.
In the Logit model the binary response
  $Y_1,\ldots,Y_n$ are assumed to be independent Bernoulli random variables with conditional mean 
  \[
  \bb E_\X [Y_i]= \frac{1}{1+\exp(-\beta_0-\v x_i^\top\v\beta)},\quad i=1,\ldots,n,
  \]
  yielding the cross-entropy loss:
\begin{equation}
    r(\v{\breve{b}})=\frac{1}{n}\sum_{i=1}^n(1-y_i)\v{\breve{x}}_i^\top\v{\breve{b}}+\ln(1+\exp(-\v{\breve{x}}_i^{\top}\v{\breve{b}})).
\end{equation}
Then, the gradient $\nabla r(\breve{\v b}_t)$ and Hessian $\m H(\breve{\v b}_t)$ are given by the formulas:
 \[
 \begin{split}
 	\mu_{t,i}&:= (1+\exp(-\breve{\v x}_i^\top\breve{\v b}_t))^{-1},\\
 	\v\mu_t&:=[\mu_{t,1},\ldots, \mu_{t,n}]^\top,\\
 	\nabla r(\breve{\v b}_t)&=\frac{1}{n} \breve{\X}^\top(\v\mu_t-\v y),\\
 	\v w_t&:=\sqrt{\v \mu_t\odot (\v 1-\v\mu_t)},\qquad ( w_{t,i}=\sqrt{\mu_{t,i}(1-\mu_{t,i})},\;\forall i),\\
 	\m D_t&:=\mathrm{diag}(\v w_t),\\
 	\m H(\breve{\v b}_t)&=\frac{1}{n}\breve{\X}^\top\m D_t^2\breve{\X}=\frac{1}{n}\sum_{i=1}^n w_{t,i}^2\breve{\v x}_i\breve{\v x}_i^\top.
 \end{split}
 \]
 If we define the quantities:
  \[
 \begin{split}
 	\X_t&:=\m D_t\X, \\
 	\v y_t&:=\v w_t b_{t,0}+\X_t\v b_t+\m D^{-1}_t(\v y-\v\mu_t),
 \end{split}
 \]
  then straightforward algebra shows that the estimator in \eqref{Taylor} can also be written in terms of the following (weighed) regularized least-squares:
 \[
 (b_{t+1,0},\v b_{t+1}):=\argmin_{b_0,\v b} \|\v y_t-\v w_t b_0-\X_t\v b\|^2+\lambda\sum_{i=1}^p\upupsilon_i\times |b_i|.
 \]
We can  eliminate the intercept term $b_0$ from the optimization by  applying the  centering (projection) matrix
 \[
 \m C_t:=\m I_n-\v w_t\v w_t^\top/\|\v w_t\|^2
 \]
 to both $\X_t$ and $\v y_t$. Once the $\v b_{t+1}$ is computed, then
 we recover $b_{0,t+1}=\v w_t^\top(\v y_t-\X_t\v b_{t+1})/\|\v w_t\|^2$.
 This gives the following formulas for updating  $(b_{0,t},\v b_{t})$ to $(b_{0,t+1},\v b_{t+1})$:
 \[
 \begin{split}
 	\v b_{t+1}&:=\argmin_{\v b}\| \m C_t\v y_t-\m C_t\X_t\v b\|^2+\lambda\sum_{i=1}^p\upupsilon_i\times |b_i|\\
 	b_{0,t+1}&:=\frac{\v w_t^\top(\v y_t-\X_t\v b_{t+1})}{\|\v w_t\|^2}.
 \end{split}
 \]
 We iterate for  $t=1,2,\ldots$ until a convergence criterion is met.
 This iterative reweighed penalized least squares  is summarized in the following pseudo-algorithm \cite{friedman2010regularization}, which assumes that we compute the Lasso  estimate $	(\hat{\beta}_{0,\lambda},\hat{\v\beta}_{\lambda})$ on a grid of $m$  values (with $m=200$ being typical values): 
\[
\lambda_1<\lambda_2<\cdots<\lambda_m,
\]
 where $\lambda_m$ is  large enough so that $\hat{\v\beta}_{\lambda_m}=\v 0$; see \cite{friedman2010regularization}.
 
 	\begin{algorithm}[H]
 	\SetAlgoSkip{}
 	\DontPrintSemicolon
 	\SetKwInput{Input}{~input}\SetKwInOut{Output}{~output}
 	\Input{$\v y,\X$, error tolerance $\epsilon>0$ and grid  $\lambda_1<\lambda_2<\cdots<\lambda_m.$}
 	\Output{Regularized solution $(b_{0,j},\v b_j)$ for each $\lambda_j\,\;j=1,\ldots,m$} 
 	$\v b\leftarrow \v 0,\;b_0\leftarrow -\ln(1/\bar y-1)$  	\tcp*[f]{\scriptsize initializing values}\;
 	\For(\tcp*[f]{\scriptsize outer loop over $\{\lambda_j\}$}){$j=m,m-1,m-2,\ldots,2,1$ }{
 		$t\leftarrow 0$\;
 		\Repeat(\tcp*[f]{\scriptsize middle loop is over the quadratic approximation}){$\|\v b_\mathrm{old}-\v b\|<\epsilon$}{
 			$\v b_\mathrm{old}\leftarrow\v b$\;
 			$\v \mu\leftarrow (\v 1+\exp(-\v 1b_0-\X\v b))^{-1}$ \tcp*[f]{\scriptsize mean response}\;
 			$\v w\leftarrow \sqrt{\v\mu\odot(\v 1-\v \mu)}$ \tcp*[f]{\scriptsize weights}\;
 			$\X_t\leftarrow \mathrm{diag}(\v w) \X$ \label{alg:weights goto}\;
 			$\v y_t\leftarrow b_0\v w+\X_t\v b+ (\v y-\v \mu)\div\v w$ \tcp*[f]{\scriptsize adjusted \& weighted response}\;
 			$c_y\leftarrow \v y_t^\top\v w/\|\v w\|^2$,\quad $\v c_x\leftarrow \X_t^\top\v w/\|\v w\|^2$ \tcp*[f]{\scriptsize adjustments for intercept}\;
 			 	$\upupsilon_i\leftarrow \sqrt{\|\m C\v v_i\|^2/n}$ for $i=1,\ldots,p$ \tcp*[f]{\scriptsize square root of feature variance}\label{alg:bo goto}\;
 			$\v b\leftarrow \mathrm{lassoCD}(\v b,\v y_t-\v w c_y,\X_t-\v w \v c_x^\top,\epsilon,\lambda_j,\v\upupsilon)$ 	\tcp*[f]{\scriptsize inner loop}\label{alg:lasso-Logistic goto}\;
 			$b_0\leftarrow c_y-\v c_x^\top\v b$ \tcp*[f]{\scriptsize intercept update given $\v b$}\;
 			$t\leftarrow t+1$\;
 		}
 		$(b_{0j},\v b_j)\leftarrow (b_0,\v b)$ \tcp*[f]{\scriptsize store values for each grid point}}
 	\KwRet{$b_{0j},\v b_j,j=1,\ldots,m$}
 	\caption{IRLS for Lasso-Logistic model \cite{friedman2010regularization}.}
 	\label{alg:lasso-Logistic}
 \end{algorithm}
 
 \medskip
 
Line \ref{alg:lasso-Logistic goto} in Algorithm~\ref{alg:lasso-Logistic} calls the subroutine $\mathrm{lassoCD}(\v b,\v y,\X,\epsilon,\lambda,\v\upupsilon)$ to compute the Lasso estimate:
\begin{equation}
	\label{simplest lasso}
\argmin_{\v b} \frac{\|\v y-\X\v b\|^2}{2n}+\lambda\sum_{i=1}^p\upupsilon_i\times |b_i|
\end{equation}
 to within an error tolerance of $\epsilon$ via the method of coordinate descent.  For completeness, we include the pseudocode of this subroutine in the Appendix and refer the reader to \cite{friedman2010regularization} for more details of this well-known algorithm.

 We now describe our proposed method for approximately computing the Lasso estimate \eqref{glm new}.  The basic idea is to modify the standard linearization \eqref{Taylor} by replacing  the penalty term $\sum_{i=1}^p\upupsilon_i\times |b_i|$ with  $\sum_{i=1}^p \upupsilon_{t,i}\times |b_i|$, where $\v\upupsilon_t$ is determined from the columns of the Hessian of the cross-entropy loss evaluated at the current  estimate $\breve{\v b}_t$. This gives an iterative reweighed least squares in which the scaling  in the lasso penalty term is modified from one iteration to the next:

 \begin{equation*}
	\breve{\v b}_{t+1}:=\argmin_{\breve{\v b}}\;(\breve{\v b}-\breve{\v b}_t)^\top\nabla r(\breve{\v b}_t)+\frac{ (\breve{\v b}-\breve{\v b}_t)^\top\m H(\breve{\v b}_t)(\breve{\v b}-\breve{\v b}_t)}{2}+\lambda\sum_{i=1}^p \upupsilon_{t,i}\times|b_i|.
\end{equation*}
We implement this iterative rescaling with only one minor modification of Algorithm~\ref{alg:lasso-Logistic}. In particular, we replace line~\ref{alg:bo goto} in Algorithm~\ref{alg:lasso-Logistic} with the  variances of the features in $\X_t$ weighted by $\sqrt{\mu_{t,i}(1-\mu_{t,i})},\;\forall i$:
\[
\v\upupsilon_t\leftarrow \sqrt{\mathrm{diag}((\X_t-\v w \v c_x^\top)^\top(\X_t-\v w \v c_x^\top))/n}.
\]
In other words, rather than computing the square root of the diagonal elements of the matrix $\X^\top\m C\X/n$ (as currently done in line~\ref{alg:bo goto} of Algorithm~\ref{alg:lasso-Logistic}), we instead compute the square root of the diagonal elements of the matrix $\X_t^\top\m C_t\X_t/n$. This is the only significant difference between the implementations of the current widely-used constant scaling and our proposed iteratively rescaled Lasso (IRL) algorithm. 

\begin{remark}[An Alternative Computation via a Pilot Estimate.]
Recall that the adaptive Lasso estimator \cite{zou2006adaptive} is a variation of the classical or vanilla Lasso estimator \eqref{lasso}, where the scaling is not constant, but depends on a root-$n$ consistent pilot estimator $\hat{\v\beta}$ of $\v\beta$ (for example, the least-squares estimator when $n>p$):
\[
\upupsilon_k(\hat{\v\beta})=\frac{\|\m C\v v_i\|^2/n}{|\hat\beta_k|}.
\]
 This type of scaling successfully achieves model selection consistency  under weak assumptions; see \cite{zou2006adaptive}. 

As in the adaptive Lasso case, given a good pilot estimate $\hat{\v\beta}$ of $\v\beta$, we can 
compute $\upupsilon_k^2(\hat{\v\beta}),\;k=1,\ldots,p$  as the diagonal elements of the Hessian matrix of the cross-entropy loss at  $\hat{\v\beta}$ and then deliver the Lasso estimator:
\begin{equation*}
	\argmin_{\v b,b_0}r(b_0,\v b)+\lambda\sum_{i=1}^p\upupsilon_i(\hat{\v\beta})\times|b_i|.
\end{equation*}
 In this implementation, the scaling parameter $\upupsilon_i(\hat{\v\beta}),\;i=1,\ldots,p$ are fixed at the beginning of the lasso optimization and do not change during the course of the iterative reweighed least squares. This approach is an alternative to the iteratively rescaled Lasso described above, which we do not pursue further in this work. 
 \end{remark}
 
\section{Numerical Results}
\label{sec: numerical}
In this section, we provide empirical comparisons between the IRL method and the state-of-the-art Glmnet implementation \cite{friedman2010regularization} for Logistic and Poisson regression. We provide numerical results for experiments conducted on both simulated and real data.

\subsection{Simulations}
In our simulations, we generate synthetic data from the model: 
\begin{equation}
    \label{eq: data_gen}
    \v h = \tau\left(\v1\beta_0 +\m \X \v \beta  \right), 
\end{equation}
where the parameter $\tau>0$ is a scaling parameter that determines the strength of the signal.
When implementing Logistic regression, we sample a random vector $\v Y$ by drawing components $y_i$ from Bernoulli($\mu_i$), where $\mu_i = 1/(1+\exp(-h_i))$ for all $i$. Similarly, for Poisson regression we sample $\v Y$ by drawing $y_i$ from Poisson$(\mu_i)$, where $\mu_i = \exp(h_i)$. Each row of the predictor matrix $\m \X$ is generated from a multivariate normal distribution with zero mean and covariance $\m \Sigma$ with diagonal elements $\m \Sigma_{j,j} = 1$ and off-diagonal elements $\m \Sigma_{i,j} = \rho^{\gamma\mid i-j\mid}$, for parameters $\rho\in(-1,1)$ and $\gamma>0$; see, e.g.,  \cite{hastie2020best,bertsimas2016best}. To induce sparsity in $\m \X$ we introduce the parameter $\xi>0$  and set $\m \X_{i,j} = 0$ for all $\m \X_{i,j} < \xi$. We let the intercept $\beta_0 = 0$ and set the first 10 components of $\beta$ as
\[
\v \beta_{[1:10]} = \begin{bmatrix} 
25,&4,& -4,&50,&4,&-4,&75,&4,&-4,& 100
\end{bmatrix}^{\top},
\]
and the remaining components to 0.

We run IRL and compare its statistical performance against the state-of-the-art implementation for generalized linear Lasso regression, Glmnet \cite{friedman2010regularization}. To tune the parameter $\lambda$ we generate an independent validation set from the generating process \eqref{eq: data_gen} with identical parameter values for $\tau$, $\xi$, $\rho$ and $\gamma$. We then minimize the expected cross-entropy error on the validation set over a grid with 100 values.

Since the vector $\v\mu=\bb{E}_{\X}[\v Y]$ is known, when implementing Logistic regression, we can use the conditional expected cross-entropy error,
\begin{equation}
    \label{eqn: pred_error_Logistic}
    \bb{E}_{\X}\lt[r(\v{\breve{b}})\rt]=\frac{1}{n}\sum_{i=1}^n(1-\mu_i)\v{\breve{x}}_i^\top\v{\breve{b}}+\ln(1+\exp(-\v{\breve{x}}_i^{\top}\v{\breve{b}})).
\end{equation}
For Poisson regression, we use the following conditional expected cross-entropy error,
\begin{equation}
    \label{eqn: pred_error_poiss}
    \bb{E}_{\X}\lt[r(\v{\breve{b}})\rt]=\frac{1}{n}\sum_{i=1}^n\exp(\v{\breve{x}}_i^\top\v{\breve{b}})-\mu_i\v{\breve{x}}_i^{\top}\v{\breve{b}}.
\end{equation}
The expected loss conditioned on $\X$ allows one to estimate the true expected generalization risk.  

 In each simulation, after generating a training and validation set ($n = 1000$, $p = 100$) we run IRL and Glmnet to evaluate the $\lambda$ that minimizes either \eqref{eqn: pred_error_Logistic} or \eqref{eqn: pred_error_poiss} on the validation set. We denote this minimizer as $\lambda^*$ and the corresponding model coefficient estimate as $\hat{\v \beta}_{\lambda^*}$. We measure the following:  
\begin{enumerate}
    \item \textbf{Bias}: This is defined as $\|\v\beta- \hat{\v \beta}_{\lambda^*}\|_2$.
    \item \textbf{True Positives (TP)}: The number of correctly identified non-zero features in $\hat{\v \beta}_{\lambda^*}$.
    \item \textbf{False Positives (FP)}: The number of falsely identified non-zero features in $\hat{\v \beta}_{\lambda^*}$.
    \item \textbf{Test Loss}: The error \eqref{eqn: pred_error_Logistic} (Logistic) or \eqref{eqn: pred_error_poiss} (Poisson) evaluated on an independently generated test set from model \eqref{eq: data_gen} of size $n$.
\end{enumerate}

For each value of the pair ($\gamma$, $\rho$) we replicate the simulation 100 times and report the mean value of each performance measure in Tables ~\ref{tbl:1} to ~\ref{tbl:6}. Standard errors of the mean are provided in parentheses.

\begin{table}[H]
  \begin{center}
  \caption{\textbf{Logistic} - $\X$ is sparse ($\xi = 0.1$, $\tau = 1$).}
  \label{tbl:1}
    \begin{tabularx}{\textwidth}{lcCCCC}
        \toprule
        \multicolumn{1}{c}{} & \multicolumn{1}{c}{} & \multicolumn{4}{c}{Average Variance = 0.047}  \\
        \cmidrule(rl){3-6} 
          Method         & $(\rho, \gamma)$       & Bias        & TP             & FP             & Test Loss                  \\
        \cmidrule(r){1-1} \cmidrule(l){2-2}\cmidrule(rl){3-6} 

        IRL & (0.1,0.1) & 73.72 & 8.58 (0.13) & 19.37 (0.43) & 0.12 \\ 
        Glmnet       &           & 125.31 & 7.28 (0.11) & 48.87 (0.74) & 0.17 \\ 
                     &           &              &            &              &          \\ 
        IRL & (0.1,1.0) & 89.73 & 9.97 (0.02) & 24.71 (0.53) & 0.04 \\ 
        Glmnet       &           & 126.59 & 9.86 (0.04) & 47.45 (0.71) & 0.08 \\ 
                     &           &              &            &              &          \\ 
        IRL & (0.9,0.1) & 61.09 & 6.65 (0.13) & 8.12 (0.29) & 0.29 \\ 
        Glmnet       &           & 111.46 & 4.84 (0.08) & 36.36 (0.68) & 0.32 \\ 
                     &           &              &            &              &          \\ 
        IRL & (0.9,1.0) & 67.47 & 7.49 (0.13) & 14.04 (0.38) & 0.18 \\ 
        Glmnet       &           & 121.52 & 6.33 (0.12) & 50.73 (0.73) & 0.22 \\ 
                     &           &              &            &              &          \\
 \bottomrule
    \end{tabularx}
    \end{center}
\end{table}

\begin{table}[h]
  \begin{center}
  \caption{\textbf{Logistic} - $\X$ is not sparse ($\xi = \infty$, $\tau = 0.01$).}
  \label{tbl:4}
    \begin{tabularx}{\textwidth}{lcCCCC}
        \toprule
        \multicolumn{1}{c}{} & \multicolumn{1}{c}{} & \multicolumn{4}{c}{Average Variance = 0.156}  \\
        \cmidrule(rl){3-6} 
          Method         & $(\rho, \gamma)$       & Bias        & TP             & FP             & Test Loss                  \\
        \cmidrule(r){1-1} \cmidrule(l){2-2}\cmidrule(rl){3-6} 
        IRL & (0.1,0.1) & 40.89 & 6.09 (0.12) & 8.30 (0.26) & 0.48 \\ 
        Glmnet       &           & 42.52 & 6.16 (0.12) & 9.65 (0.26) & 0.48 \\ 
                     &           &              &            &              &          \\ 
        IRL & (0.1,1.0) & 36.46 & 5.40 (0.10) & 14.96 (0.42) & 0.56 \\ 
        Glmnet       &           & 37.31 & 5.45 (0.11) & 15.8 (0.43) & 0.56 \\ 
                     &           &              &            &              &          \\ 
        IRL & (0.9,0.1) & 123.43 & 4.39 (0.11) & 3.09 (0.18) & 0.41 \\ 
        Glmnet       &           & 122.15 & 4.42 (0.11) & 3.85 (0.20) & 0.41 \\ 
                     &           &              &            &              &          \\ 
        IRL & (0.9,1.0) & 57.78 & 6.16 (0.13) & 5.62 (0.22) & 0.45 \\ 
        Glmnet       &           & 59.30 & 6.29 (0.12) & 7.13 (0.24) & 0.45 \\ 
                     &           &              &            &              &          \\        \bottomrule
    \end{tabularx}
    \end{center}
\end{table}

When conducting experiments with generalized linear models it is useful to estimate the noise inherent in the model. In Logistic regression, the variance of $y_i$ is given by 
\[
\var[y_i] = \mu_i(1-\mu_i).
\]
To quantify the signal strength we report the value 
\[
\sum_{i=1}^{n}\mu_i(1-\mu_i)/n
\]
averaged over all the simulated responses in each table. The average variance has a maximum value of 0.25 which indicates high noise and a weak signal whereas an average variance closer to the minimum value of 0 indicates low noise and a strong signal.

In Table ~\ref{tbl:1} we observe simulations with a strong signal. It is in this scenario that the IRL method significantly outperforms Glmnet. With IRL, we are able to obtain a sparser estimate $\hat{\v \beta}_{\lambda^*}$ that contains significantly fewer false positives and achieves improved test loss. IRL outperforms Glmnet the most when the correlation is highest among the predictors ($\rho = 0.9$, $\gamma = 0.1$). In Table ~\ref{tbl:4}, the simulated $\X$ is non-sparse, and $\tau$ is decreased thereby creating a weaker signal in the generated responses. In this high-noise setting, IRL achieves only marginally fewer false positives. We provide further numerical tables in the Appendix which contain results from simulations with more values of $\tau$.

\begin{figure*}[h!]
\centering
\begin{subfigure}{.49\textwidth}
  \centering
  \includegraphics[width=\linewidth]{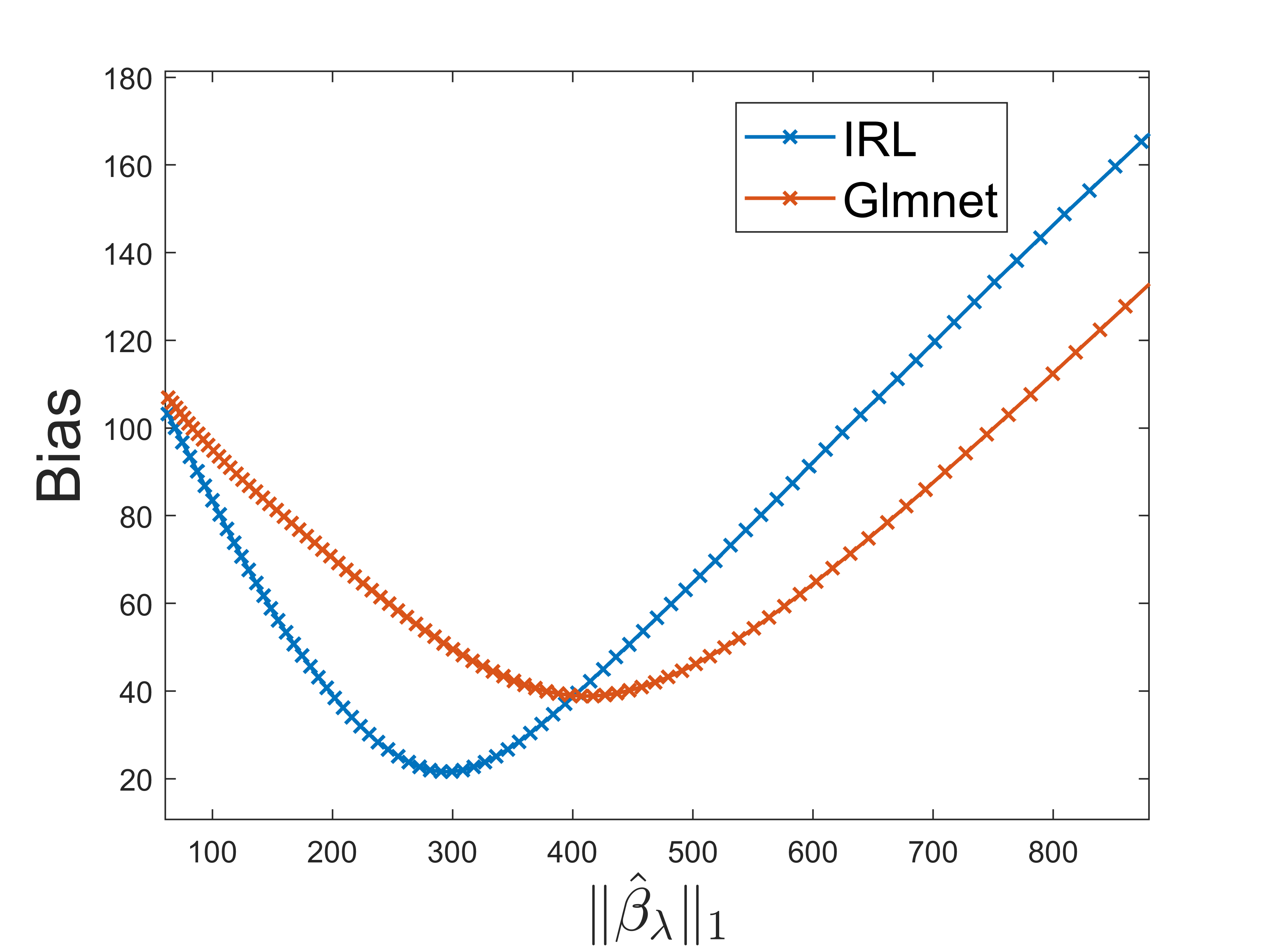}  
\end{subfigure}
\begin{subfigure}{.49\textwidth}
  \centering
  \includegraphics[width=\linewidth]{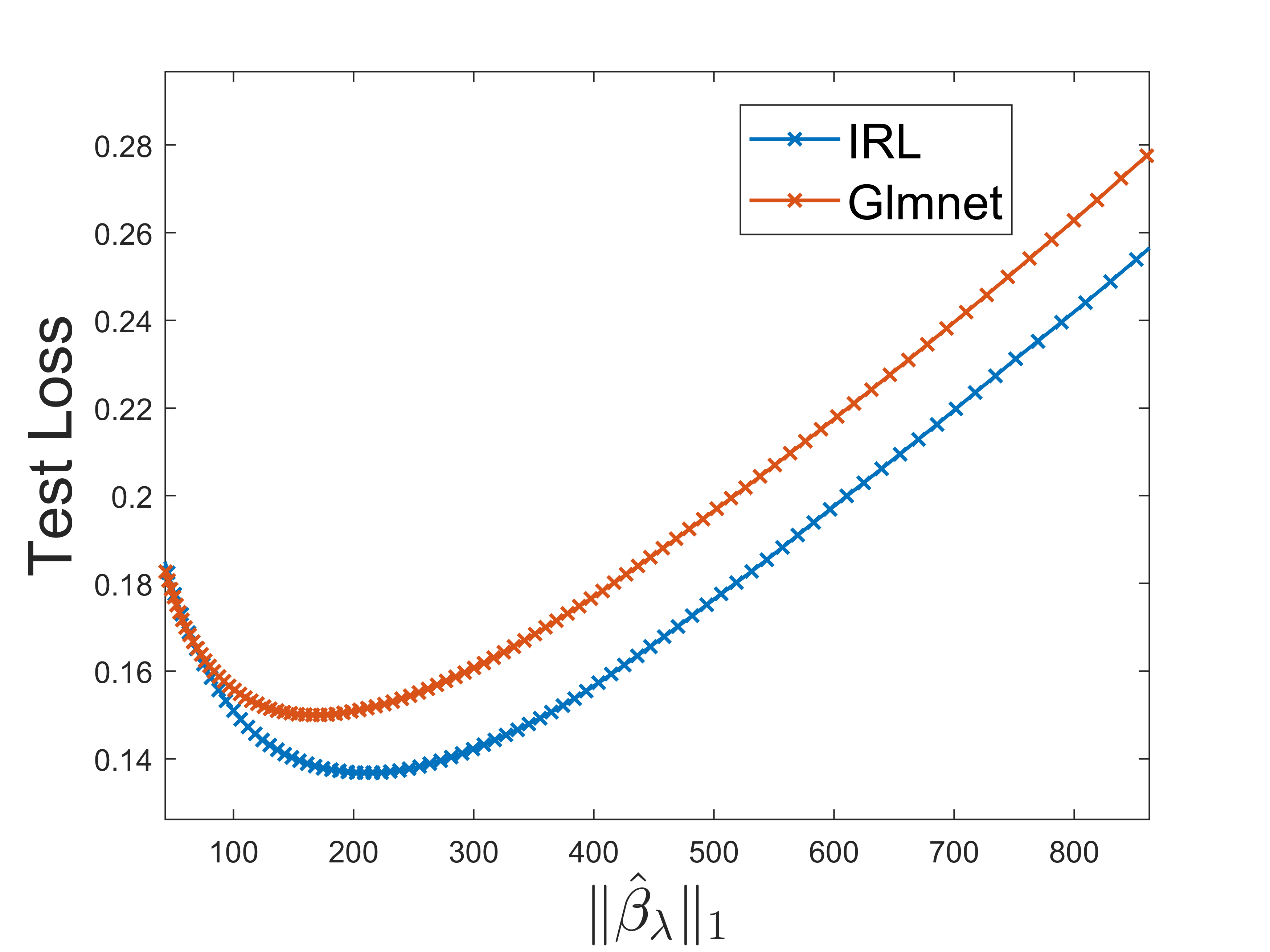}  
\end{subfigure}
\caption{Bias and test loss of $\hat{\v\beta}_{\lambda}$ in a Logistic regression simulation. The estimate $\hat{\v\beta}_{\lambda}$ is computed over a grid of 100 $\lambda$ values where the simulation parameters are $\tau = 0.01$, $\xi = 0.1$, $\rho = 0.1$ and $\gamma = 1$.}
\label{fig: logit_simulation}
\end{figure*}

In Figure ~\ref{fig: logit_simulation} we compare the bias and test loss of $\hat{\v \beta}_{\lambda}$ computed with IRL and Glmnet over a grid of $\lambda$ values. Since the parameter $\lambda$ does not coincide between IRL and Glmnet, we plot the bias and test loss as a function of the dependent variable $\|\hat{\v \beta}_{\lambda}\|_1$. Figure ~\ref{fig: logit_simulation} shows that in the simulation setting where $\X$ is sparse, IRL attains lower test loss over a range of $\lambda$ values and the lowest bias.

\begin{table}[h!]
  \begin{center}
  \caption{\textbf{Poisson} - $\X$ is sparse ($\xi = 0.1$, $\tau = 0.1$).}
  \label{tbl:5}
    \begin{tabularx}{\textwidth}{lcCCCC}
        \toprule
        \multicolumn{1}{c}{} & \multicolumn{1}{c}{} & \multicolumn{4}{c}{Average Variance = 0.289} \\
        \cmidrule(rl){3-6} 
          Method         & $(\rho, \gamma)$       & Bias        & TP             & FP             & Test Loss                  \\
        \cmidrule(r){1-1} \cmidrule(l){2-2}\cmidrule(rl){3-6} 

        IRL & (0.1,1.0) & 35.98 & 9.14 (0.09) & 20.76 (0.51) & 0.16 \\ 
        Glmnet       &           & 73.62 & 9.62 (0.06) & 47.56 (0.63) & 0.19 \\ 
                     &           &              &            &              &          \\ 
        IRL & (0.1,0.1) & 36.13 & 5.69 (0.11) & 11.03 (0.33) & 0.28 \\ 
        Glmnet       &           & 71.19 & 6.30 (0.12) & 43.78 (0.58) & 0.31 \\ 
                     &           &              &            &              &          \\ 
        IRL & (0.9,1.0) & 35.65 & 5.80 (0.11) & 8.66 (0.27) & 0.36 \\ 
        Glmnet       &           & 69.55 & 6.09 (0.12) & 41.54 (0.54) & 0.39 \\ 
                     &           &              &            &              &          \\      
        IRL & (0.9,0.1) & 41.86 & 6.09 (0.13) & 4.80 (0.23) & 0.49 \\ 
        Glmnet       &           & 61.78 & 5.36 (0.10) & 23.26 (0.42) & 0.51 \\ 
                     &           &              &            &              &          \\ \bottomrule
    \end{tabularx}
    \end{center}
\end{table}

\begin{table}[h]
  \begin{center}
  \caption{\textbf{Poisson} - $\X$ is not sparse ($\xi = \infty$, $\tau = 0.01$).}
  \label{tbl:6}
    \begin{tabularx}{\textwidth}{lcCCCC}
        \toprule
        \multicolumn{1}{c}{} & \multicolumn{1}{c}{} & \multicolumn{4}{c}{Average Variance = 8.808} \\
        \cmidrule(rl){3-6} 
          Method         & $(\rho, \gamma)$       & Bias        & TP             & FP             & Test Loss                  \\
        \cmidrule(r){1-1} \cmidrule(l){2-2}\cmidrule(rl){3-6} 

        IRL & (0.1,1.0) & 11.6 & 8.22 (0.13) & 19.02 (0.66) & -2.24 \\ 
        Glmnet       &           & 11.62 & 8.21 (0.13) & 19.37 (0.66) & -2.24 \\ 
                     &           &              &            &              &          \\ 
        IRL & (0.1,0.1) & 11.64 & 6.44 (0.09) & 8.14 (0.47) & -14.77 \\ 
        Glmnet       &           & 11.73 & 6.43 (0.10) & 8.95 (0.49) & -14.77 \\ 
                     &           &              &            &              &          \\ 
        IRL & (0.9,1.0) & 11.98 & 6.58 (0.10) & 6.72 (0.39) & -36.17 \\ 
        Glmnet       &           & 12.18 & 6.54 (0.10) & 8.30 (0.45) & -36.17 \\ 
                     &           &              &            &              &          \\ 
        IRL & (0.9,0.1) & 19.91 & 6.32 (0.13) & 2.81 (0.18) & -102.07 \\ 
        Glmnet       &           & 20.13 & 6.45 (0.13) & 4.49 (0.30) & -102.07 \\ 
                     &           &              &            &              &          \\ 
             \bottomrule
    \end{tabularx}
    \end{center}
\end{table}

\begin{figure*}[h!]
\centering
\begin{subfigure}{.49\textwidth}
  \centering
  \includegraphics[width=\linewidth]{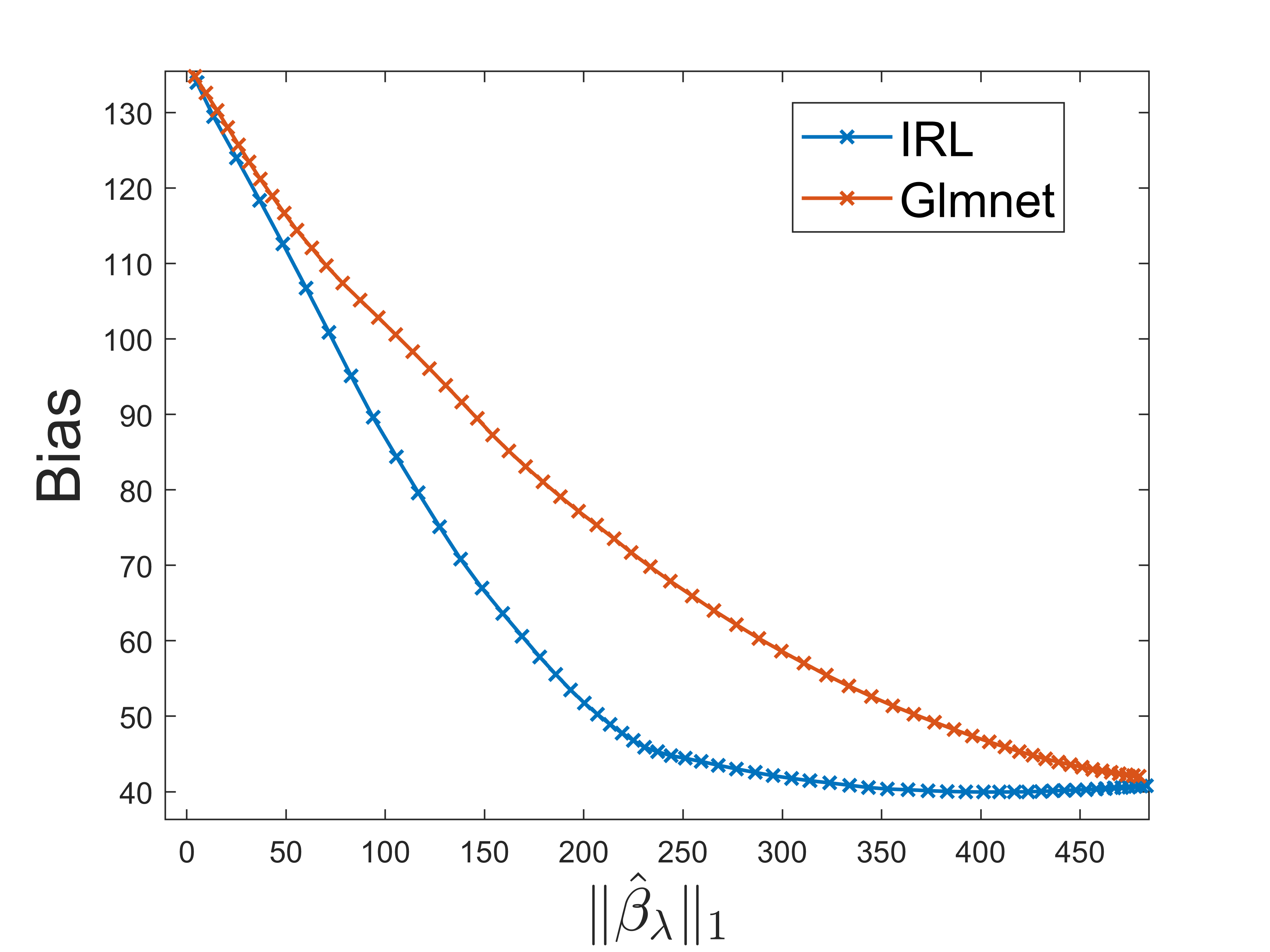}  
\end{subfigure}
\begin{subfigure}{.49\textwidth}
  \centering
  \includegraphics[width=\linewidth]{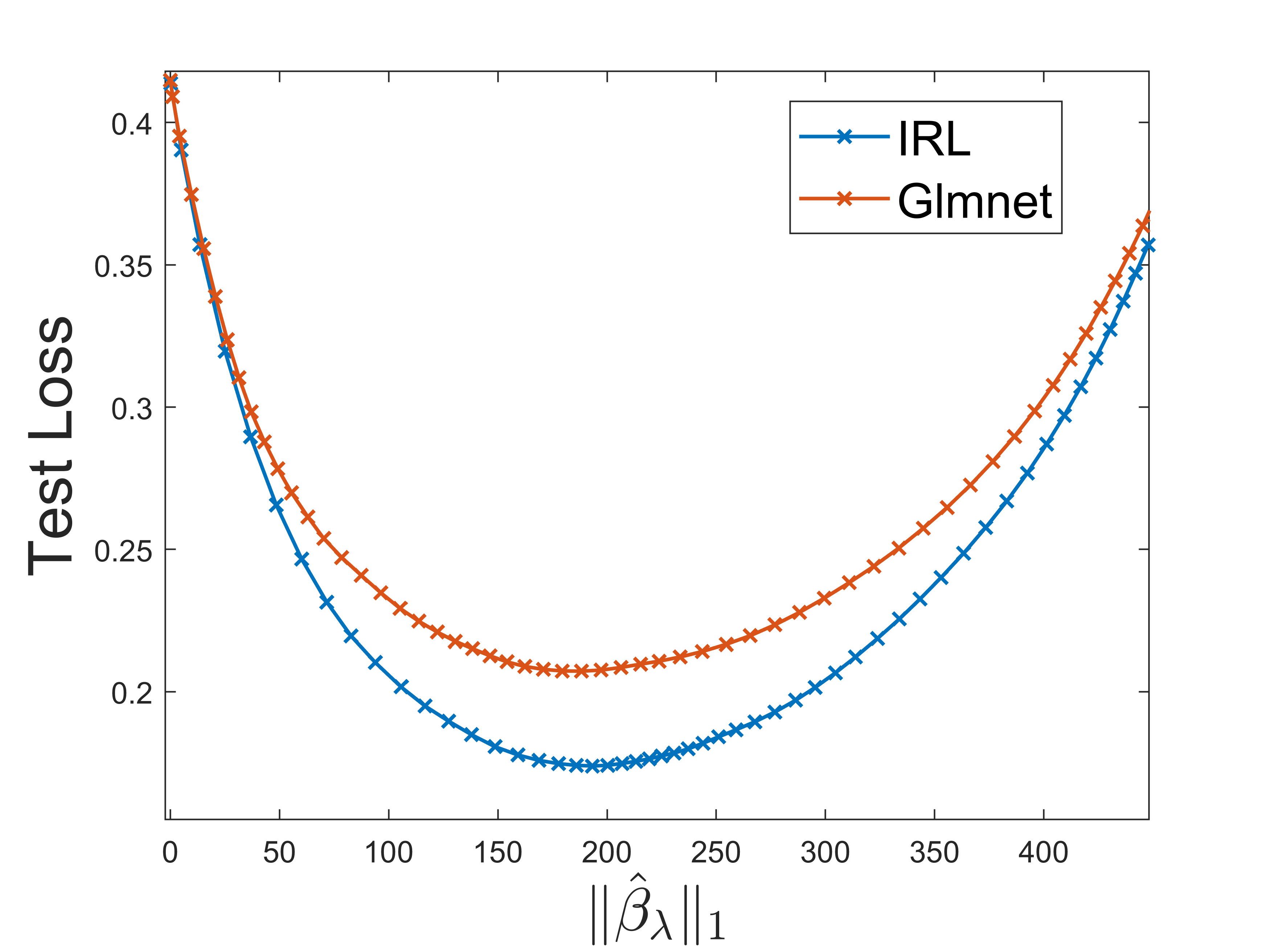}  
\end{subfigure}
\caption{Bias and test loss of $\hat{\v\beta}_{\lambda}$ in a Poisson regression simulation. The estimate $\hat{\v\beta}_{\lambda}$ is computed over a grid of 100 $\lambda$ values where the simulation parameters are $\tau = 0.1$, $\xi = 0.1$, $\rho = 0.1$ and $\gamma = 1$.}
\label{fig: poisson_simulation}
\end{figure*}

In Poisson regression the variance of $y_i$ is equal to its mean, that is, $\var[y_i] = \mu_i$. To quantify the noise we report the value $\sum_{i=1}^{n}\mu_i/n$ averaged over all the simulated responses in each table. In Table ~\ref{tbl:5} we observe simulations with low noise. Similar to Logistic regression, it is in this scenario that the IRL method significantly outperforms Glmnet. The IRL method is able to pick up significantly fewer false positives, thus exhibiting improved variable selection. The improvement is most significant when the correlation parameters are ($\rho = 0.9$, $\gamma = 1.0$). In Table ~\ref{tbl:6}, the simulated $\X$ is non-sparse, and $\tau$ is decreased to 0.01, thereby increasing the average noise to 8.08. In this scenario, IRL outperforms Glmnet only in the high correlation settings. Figure ~\ref{fig: poisson_simulation} shows that in the sparse setting, IRL outperforms Glmnet over a range of  $\lambda$ values. 

\subsection{Real Data}

\begin{figure*}[h]
\centering
\includegraphics[width=10cm]{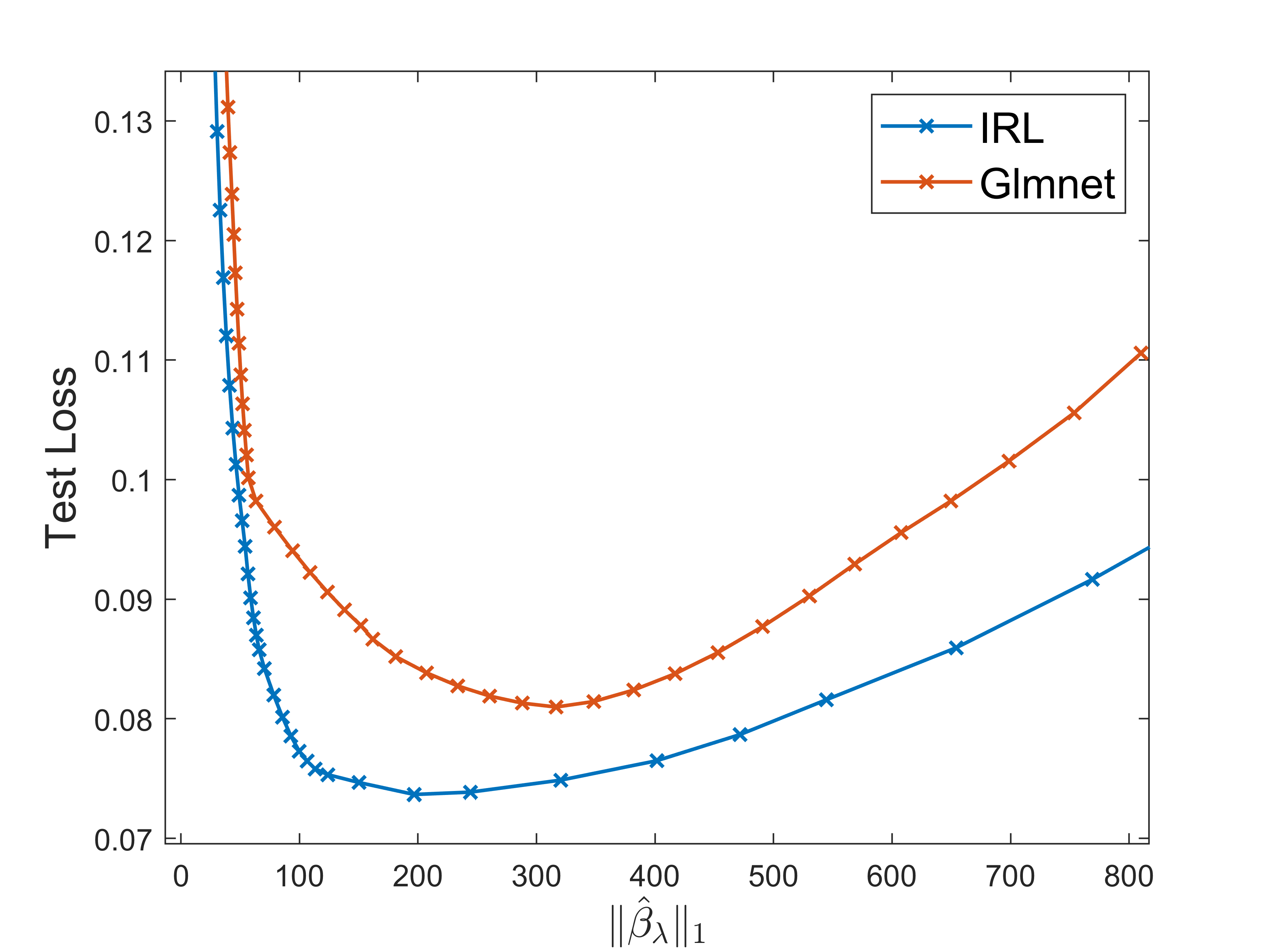}  
\caption{Test loss on the Breast Cancer Wisconsin (Diagnostic) dataset ($n = 569$, $p = 170$).}
\label{fig: real_data}
\end{figure*}

We study the performance of Glmnet and IRL on the UCI ML Breast Cancer Wisconsin (Diagnostic) dataset \cite{misc_breast_cancer_wisconsin_(diagnostic)_17}. The data set consists of 569 observations and 30 predictors. The ordering of the 569 observations is randomized and we use 399 and 170 observations for training and testing respectively. On this dataset, we fit Glmnet and IRL over a range of 100 $\lambda$ values and plot the test loss \eqref{eqn: pred_error_Logistic} in Figure \ref{fig: real_data}. The results indicate that IRL achieves the lowest test loss and this occurs when $\| \hat{\v\beta}_{\lambda}\|_1 = 223.58 $, whereas for Glmnet the minimum test loss occurs when $\|\hat{\v\beta}_{\lambda}\|_1 = 346.44$. Moreover, at their corresponding minimizers, IRL selects 13 non-zero features whereas Glmnet selects 14. In other words, the IRL can achieve a smaller test loss using a sparser model. 

In other numerical experiments with real-world data (not shown here), we similarly observed that the statistical performance of the IRL estimator, whilst  never inferior to that of the  classical/vanilla Lasso estimator,  can achieve a smaller test loss using a sparser model 
in  problems with feature structure like the ones considered in the examples above. 

\section{Concluding Remarks}

In this paper, we proposed a novel scaling of the features for Lasso penalized nonlinear regression. The current widely-used standardization of the feature matrix is performed only once, typically  as a preprocessing step, and is independent of the true regression coefficient $\v\beta$. In contrast, our rescaling proposal uses a standardization 
which depends on the current best estimate of $\v\beta$ and is thus updated from one iteration to the next during the course of the nonlinear optimization. 

Numerical experiments suggest that the proposed iteratively rescaled Lasso estimator, while never harming the model-selection performance of the estimator in any of the examples considered, has the potential to significantly improve the bias and model selection performance of the estimator when the features matrix is sparse and the signal is strong. In addition, the proposed estimator  is just as easy to compute as the vanilla Lasso estimator and only requires a minor modification of  one line of the original algorithm.

Future areas of research include  providing a theoretical framework of precise conditions  under which a significant improvement in the statistical accuracy of the IRL estimator is guaranteed. 

\appendix

\section{Coordinate Descent for Linear Lasso Computations}

The following code provides a simple coordinate descent algorithm \cite{friedman2010regularization} for computing
\eqref{simplest lasso}. Note the absence here of the intercept term $\beta_0$, which is assumed to have been eliminated from the optimization. In the pseudocode below we use the Lasso shrinkage operator, defined as:
\[
S_{\lambda}^{\mathrm{Lasso}}(x):=x[1-\lambda/|x|]_+,
\]
where $x_+=\max\{0,x\}$.

\begin{algorithm}[H]
	\SetAlgoSkip{}
	\DontPrintSemicolon
	\SetKwInput{Input}{~input}\SetKwInOut{Output}{~output}
	\Input{initial  $\v b$, scaling parameter $\v\upupsilon$,  and $\X,\epsilon,\lambda$}
	\Output{(global) minimizer $\v b$}
	$\v r\leftarrow \v y-\X\v b$ \tcp*[f]{\scriptsize initial residual}\;
	\Repeat(\tcp*[f]{\scriptsize iterate over CD cycles}){$\|\v b-\v b_\mathrm{old}\|<\epsilon$}{
		$\v b_\mathrm{old}\leftarrow \v b$\;
		\For(\tcp*[f]{\scriptsize coordinate-descent cycle}){$k=1,\ldots,p$}{
			$b_\mathrm{new}\leftarrow S_{\lambda/\upupsilon_k}^{\mathrm{Lasso}}(b_k+\v v_k^\top\v r/\|\v v_k\|^2) $ \tcp*[f]{\scriptsize this step costs $\c O(n)$}\label{alg:lasso w}\;
			\If(\tcp*[f]{\scriptsize update only if necessary}){$b_\mathrm{new}\not =b_k$}{
			$\v r\leftarrow \v r+(b_k-b_\mathrm{new})\v v_k$   \tcp*[f]{\scriptsize this update costs $\c O(n)$}\;
			$b_k\leftarrow b_\mathrm{new}$}
		     }
	}
	\KwRet{$\v b$}
	\caption{Coordinate Descent for $\frac{1}{2n}\|\v y-\X\v b\|^2+\lambda\sum_{i}\upupsilon_i\times|b_i|$}
\end{algorithm}

\section{Additional Numerical Results}

In Tables ~\ref{tbl: a1} and~\ref{tbl: a2} we provide results for numerical simulations with additional values of $\tau$. In Table ~\ref{tbl: a1}, $\X$ is simulated to be sparse and $\tau = 0.01$ which results in a weaker signal than that of in Table ~\ref{tbl:1}. In Table ~\ref{tbl: a2}, $\X$ is simulated to be non-sparse and $\tau = 0.1$ which results in a stronger signal than that of in Table ~\ref{tbl:4}. In comparison to the weak signal setting in Table ~\ref{tbl:4}, there is a greater difference in the mean number of false positives in favor of the IRL method in Table ~\ref{tbl: a2}.

\begin{table}[H]
  \begin{center}
  \caption{\textbf{Logistic} - $\m X$ is sparse ($\xi = 0.1$, $\tau = 0.01$).}
  \label{tbl: a1}
    \begin{tabularx}{\textwidth}{lcCCCC}
        \toprule
        \multicolumn{1}{c}{} & \multicolumn{1}{c}{} & \multicolumn{4}{c}{Average Variance = 0.191} \\
        \cmidrule(rl){3-6} 
          Method         & $(\rho, \gamma)$       & Bias        & TP             & FP             & Test Loss                  \\
        \cmidrule(r){1-1} \cmidrule(l){2-2}\cmidrule(rl){3-6} 
        IRL & (0.9,0.1) & 146.29 & 3.55 (0.13) & 3.61 (0.18) & 0.55 \\ 
        Glmnet       &           & 143.38 & 3.66 (0.12) & 4.56 (0.20) & 0.55 \\ 
                     &           &              &            &              &          \\ 
        IRL & (0.1,1.0) & 55.2 & 4.49 (0.11) & 12.27 (0.39) & 0.57 \\ 
        Glmnet       &           & 58.07 & 4.64 (0.12) & 13.85 (0.39) & 0.57 \\ 
                     &           &              &            &              &          \\ 
        IRL & (0.1,0.1) & 63.74 & 5.79 (0.13) & 7.37 (0.29) & 0.56 \\ 
        Glmnet       &           & 66.26 & 5.87 (0.12) & 9.29 (0.30) & 0.56 \\ 
                     &           &              &            &              &          \\ 
        IRL & (0.9,1.0) & 80.68 & 5.62 (0.12) & 5.94 (0.25) & 0.56 \\ 
        Glmnet       &           & 81.78 & 5.72 (0.12) & 7.90 (0.28) & 0.56 \\ 
                     &           &              &            &              &          \\        \bottomrule
    \end{tabularx}
    \end{center}
\end{table}

\begin{table}[H]
  \begin{center}
  \caption{\textbf{Logistic} - $\m X$ is not sparse ($\xi = \infty$, $\tau = 0.1$ ).}
  \label{tbl: a2}
    \begin{tabularx}{\textwidth}{lcCCCC}
        \toprule
        \multicolumn{1}{c}{} & \multicolumn{1}{c}{} & \multicolumn{4}{c}{Average Variance = 0.021}  \\
        \cmidrule(rl){3-6} 
          Method         & $(\rho, \gamma)$       & Bias        & TP             & FP             & Test Loss                  \\
        \cmidrule(r){1-1} \cmidrule(l){2-2}\cmidrule(rl){3-6} 

        IRL & (0.1,0.1) & 49.29 & 6.97 (0.11) & 23.59 (0.46) & 0.08 \\ 
        Glmnet       &           & 53.75 & 6.85 (0.10) & 30.79 (0.48) & 0.09 \\ 
                     &           &              &            &              &          \\ 
        IRL & (0.1,1.0) & 48.54 & 9.34 (0.07) & 43.09 (0.54) & 0.12 \\ 
        Glmnet       &           & 50.32 & 9.40 (0.07) & 47.15 (0.54) & 0.12 \\ 
                     &           &              &            &              &          \\ 
        IRL & (0.9,0.1) & 58.08 & 6.34 (0.12) & 4.65 (0.24) & 0.06 \\ 
        Glmnet       &           & 66.95 & 6.42 (0.13) & 10.96 (0.28) & 0.06 \\ 
                     &           &              &            &              &          \\ 
        IRL & (0.9,1.0) & 45.34 & 6.77 (0.11) & 15.47 (0.37) & 0.07 \\ 
        Glmnet       &           & 53.09 & 6.53 (0.11) & 23.93 (0.42) & 0.07 \\ 
                     &           &              &            &              &          \\         \bottomrule
    \end{tabularx}
    \end{center}
\end{table}

\bibliographystyle{splncs04}
\bibliography{bibliography}

\end{document}